\documentclass[a4paper]{article}
\usepackage{authblk}
\usepackage{graphicx}
\usepackage{amsmath}
\usepackage{amssymb}
\usepackage{caption}
\usepackage{subcaption}
\begin{document}

%\title[Self-organized Escape Process in a Nonlinear
%Potential]{Self-organized escape processes of linear chains in
%  nonlinear potentials}\label{ra_ch1}

\title{Self-organized escape processes of linear chains in
  nonlinear potentials}\label{ra_ch1}
\author[1]{Torsten Gross\thanks{tgross@physik.hu-berlin.de}}
\author[2]{Dirk Hennig}
\author[1]{Lutz Schimansky-Geier}
\affil[1]{Department of Physics, Humboldt-University at Berlin, Newtonstr. 15, D-12489 Berlin, Germany}
\affil[2]{Department of Mathematics, University of Portsmouth, Portsmouth, PO1 3HF, United Kingdom}

\date{May 10, 2013}
%\author[T. Gross, D.  Hennig, and L. Schimansky-Geier]{Torsten
%  Gross\footnote{e-Mail:tgross@physik.hu-berlin.de}, Dirk Hennig, and
%  Lutz Schimansky-Geier}
%\index[aindx]{Author, F.} % or \aindx{Author, F.}
%\index[aindx]{Author, S.} % or \aindx{Author, S.}

%\address{Department of Physics, Humboldt-University at Berlin,
%  Newtronstr.15, D-12489 Berlin, Germany\\Department of Mathematics,
%  University of Portsmouth, Portsmouth,\\PO1 3HF, United Kingdom}

\maketitle

\begin{abstract}
  An enhancement of localized nonlinear modes in coupled systems gives rise to a novel type of escape process.  We study a spatially one dimensional set-up consisting of a linearly coupled oscillator chain of
  $N$ mass-points situated in a metastable nonlinear potential. The Hamilton-dynamics exhibits breather
  solutions as a result of modulational instability of the phonon
  states. These breathers localize energy by freezing other parts of
  the chain. Eventually this localised part of the chain grows in
  amplitude until it overcomes the critical elongation characterized by the
  transition state. Doing so, the breathers ignite an escape by
  pulling the remaining chain over the barrier. Even if the formation of
  singular breathers is insufficient for an escape, coalescence of
  moving breathers can result in the required concentration of energy.
  Compared to a chain system with linear damping and thermal
  fluctuations the breathers help the chain to overcome the
  barriers faster in the case of low damping.  With
  larger damping, the decreasing life time of the breathers effectively inhibits the escape process.
\end{abstract}

%\markright{Customized Running Head for Odd Page} % default is chapter title.

%\cite{Langer,Hanggi90}
%\cite{Sodano,HanMarRis,Christen95,Christen_EPL,Christen_PRE,Pankratov,Gulevich}
\section{Introduction}
\label{intro}
A chain of binary interacting units is a simple model for discussing
the emergence of collective phenomena. Despite its simplicity, such
setup appears frequently in various physical contexts such as for the
description of mechanical and electrical systems, polymers, networks
of superconducting elements, chemical reactions in connected discrete
boxes, to name but a few
\cite{Sato06_Rev,Flach98_PR,Dauxois92_PD,Reineker89_PLA,Doi,Nicolis,MakNek97}.

In this study we use the linear chain and its cooperative dynamical
phenomena as a paradigm of a multidimensional dynamical system. We aim
to investigate escape processes of the chain out of a metastable
state\cite{Langer,Hanggi90} also known as the nucleation of a
kink-antikink pair
\cite{Sodano,HanMarRis,Christen95,Christen_EPL,Christen_PRE,Pankratov,Gulevich}
in biased sinusoidal potentials. To this end we place a
chain with linear springs being responsible for the interaction
between the units in a nonlinear potential 
modelled by polynomial of 3rd degree. As will be seen, energy along
the chain will become inhomogeneously distributed and parts of the
chain with large elongations will collect energy from their
neighbouring regions. Such localized modes of energy are know as
breather-solutions and have been studied intensively in the past in
various contexts including micro-mechanical cantilever arrays
\cite{Sato06_Rev,cantilever1,cantilever2}, arrays of coupled Josephson junctions
\cite{josephson1,josephson2}, coupled optical wave guides
\cite{optical1,optical2}, Bose-Einstein condensates in optical
lattices \cite{BEC}, in coupled torsion pendula \cite{Jesus},
electrical transmission lines \cite{electrical1,electrical2}, and
granular crystals \cite{crystals}.

We concentrate here on the escape process and elaborate how the
localized breathers modify this process \cite{HeFug07,HeLsg07}. For
this purpose we consider first the pure deterministic set-up and study
the properties of breathers arising on the chain whose units evolve in
the nonlinear potential. In the second set-up we investigate thermally
activated escape dynamics. The chain will be exposed to a thermal bath
with temperature $T$. Consequently damping and noise is added to the
deterministic dynamics accounting for coupled Langevin equations.  Our
main new findings concern the study of how a change of the friction
coefficient modifies the escape process.  While for stronger damping
breather solutions do not play a significant role, in case of weak
damping the escape times become even shorter compared to the
deterministic case. Notably, the establishment of breathers along the
chain helps the emergence of critical elongations from thermal fluctuations.

This work is structured as follows: In Sec.~\ref{1d:1d-chain_model} we
introduce the model of a linearly coupled chain situated in an
external nonlinear potential describing a metastable situation.  We
study the critical transition state, i.e.  the bottleneck
configuration which the chain has to cross in order that a transition
over the potential barrier takes place. In Sec.~\ref{sec2} we derive
conditions for the modulational instability which determine the time
scale for the growth of the breathers. We find two generic scenarios
which govern the transition. With larger energy singular breathers
achieve large elongations and can surpass the transition states alone.
Differently, if the elongation of the breathers are too small, they undergo an erratic motion. On collision, breathers tend to merge. Thereby they cumulatively localize energy which can eventually cause the barrier crossing.  In Sec.~\ref{sec3}
we study the thermally activated escape of the interacting
chain in the metastable potential landscape.  Finally,
we summarize our findings.

\section{The one-dimensional chain model}\label{1d:1d-chain_model}
We study an one-dimensional chain of $N$ linearly coupled oscillators
of mass $m$ with elongations $q_n(t), n=1,\ldots,N$. The chain is
positioned in a cubic external potential. Every mass point experiences
a nonlinear force caused by the potential
\[V(q_n)=\frac{m\,\omega_0^2}{2}q^2_n-\frac{a}{3}q^3_n
\]
and spring forces created from the neighbours with spring constant
$\kappa$. Periodic boundary conditions are applied. Positions of
particles perpendicular to the potential variation are kept
constant\cite{HeLsg07}, (for an alternative case see
\cite{FugHe07,martens08}).  First, we assume that there is no noise
and no damping, hence yielding a canonic situation with a Hamiltonian
dynamics and corresponding momenta $p_n(t), n=1,\ldots,N$ canonically
conjugate to the positions $q_n(t)$.  Consequently the total energy of
the chain is conserved.

In order to obtain dimensionless quantities we rescale units and parameters,
$\widetilde{q_n}=a/(m\,\omega_0^2)\,q_n$,
${\widetilde{p}_n}\,^2=a^2/(m^4\,\omega_0^6)p_n$ and
${\widetilde{t}}\,^2=\omega_0^2\,t^2$. As a result, we remain with
dimensionless Hamiltonian with one remaining parameter only, the effective
coupling strength $\widetilde{\kappa}=\kappa/(m\,\omega_0^2)$. In what follows we omit the tildes.

The Hamiltonian of the considered chain reads:
\begin{align*}
  \mathcal{H}=\sum_{n=0}^{N-1}\left[\frac{p_n^2}{2}+\frac{\kappa}{2}\left(q_n-q_{n+1}\right)^2+V\left(q_n\right)
  \right], \qquad V(q_n)=\frac{q_n^2}{2}-\frac{q_n^3}{3}.
\end{align*}
The resulting equations of motion become
\begin{align}
  \label{equ_motion_1d}
  \ddot{q_n}+q_n-{q_n}^2-\kappa \left(q_{n+1}+q_{n-1}-2\,q_n\right)=0,
  \qquad q_{N+1}=q_1.
\end{align}
In this paper we will consider for our numerical simulations chains
comprising $N=100$ units. A study of the dependence of the escape
process on the number of oscillators can be found in
\cite{HeFug07}.

For the study of an escape, we initially place the units of the chain close to the bottom of
the external potential, that is nearby  $q_{min}=0$ and provide them with
energy $E$. As will be seen, the chain eventually generates critical
elongations surpassing the potentials local maximum. This initiates a
transition of the chain into the unbounded regime $q_n > q_{\mathrm{max}}=1,
n=1,\ldots,N$, which we refer to as an escape. For a single particle to overcome the potential barrier it needs to be supplied with an energy $\Delta E=V(1)-V(0)=1/6$.

In the following we want to illustrate that the generation of these
critical states is efficient even in cases where the chain energy $E$
is small compared to $E \ll N\cdot \Delta E$. This low-energy setting
is obtained through the following initial preparation of the system
\begin{align*}
q^{\mathrm{IC}}_i=\Delta q_i+\Delta \qquad
p^{\mathrm{IC}}_i=\Delta p_i,
\end{align*}
where $\Delta q_i$ and $\Delta p_i$ are small random perturbations
taken from a uniform distribution within the intervals
\begin{align*}
  \Delta q_i\in \left[-\Delta q^{\mbox{IC}},\Delta q^{\mbox{IC}}
  \right] \text{ and } \Delta p_i\in \left[-\Delta
    p^{\mbox{IC}},\Delta p^{\mbox{IC}} \right],
\end{align*}
and the coordinate shift, $0< \Delta <1$, is chosen to increase the
system energy to a desired value. This procedure results in a variety
of perturbed flat initial states comprising a specific energy which we
view as a statistical ensemble.

\subsection{Transition states}
For Hamiltonian systems the local minima of the energy surface (in
phase space) are Lyapunov stable. That is, orbits in the vicinity of a
local minimum never leave it as their associated energy is conserved.
Therefore, orbits with an energy exceeding the energy associated with
a neighbouring saddle point of first order are no longer bound to the
basin. Thus the saddle point is referred to as transition state, as it
separates bounded from unbounded orbits. Concerning our problem, the
system's energy has to exceed the transition state energy to make
escape events possible. To determine this transition states, we have
to solve $\nabla U(q_1,q_2,\ldots)=0$, where $U$ denotes the potential
energy (thus the transition state is a fixed-point) and the solution
must render all eigenvalues of the Hessian matrix of $U$ positive,
except for a single negative one. In general, this is a non-trivial
task requiring sophisticated numerical methods. Here, we used the dimer method. It is a minimum-mode following method that
solely makes use of gradients of the potential surface. It was first
introduced in \cite{TS:henkelman} and its computational effort scales
favorable with the system size.

In the one-dimensional chain model the transition state configurations
solve the stationary equation\footnote{An alternative approach for the
  one-dimensional chain model is presented in
  \cite{HeLsg07}. It casts the stationary
  equation into a two-dimensional map and links the localized lattice
  solutions to its homoclinic orbits.}
\begin{align}
\label{ts1d:equ_sad_point_1d}
q_n-{q_n}^2-\kappa \left(q_{n+1}+q_{n-1}-2\,q_n\right)=0
\end{align}
and fulfill the condition on the eigenvalues, $\lambda^H$, of the
Hessian matrix $H$
\[H_{i,j}=\delta_{i,j}(2\,\kappa+1-2\,q_i)-\kappa\,\left(\delta_{i,j+1}+\delta_{i,j-1}\right).  \]

\begin{figure}
\centering
\includegraphics[width=0.3\linewidth]{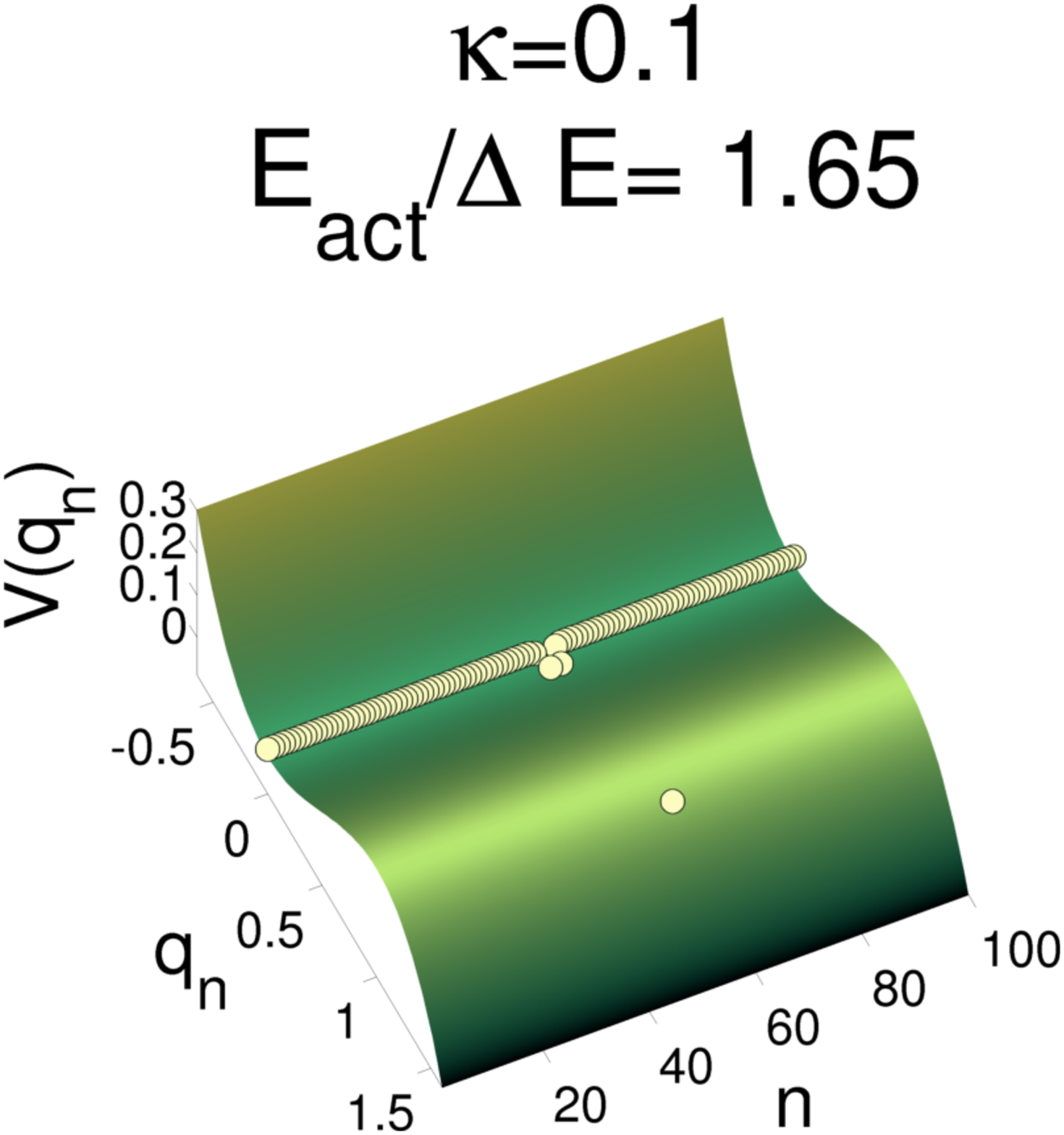}\hfill
\includegraphics[width=0.3\linewidth]{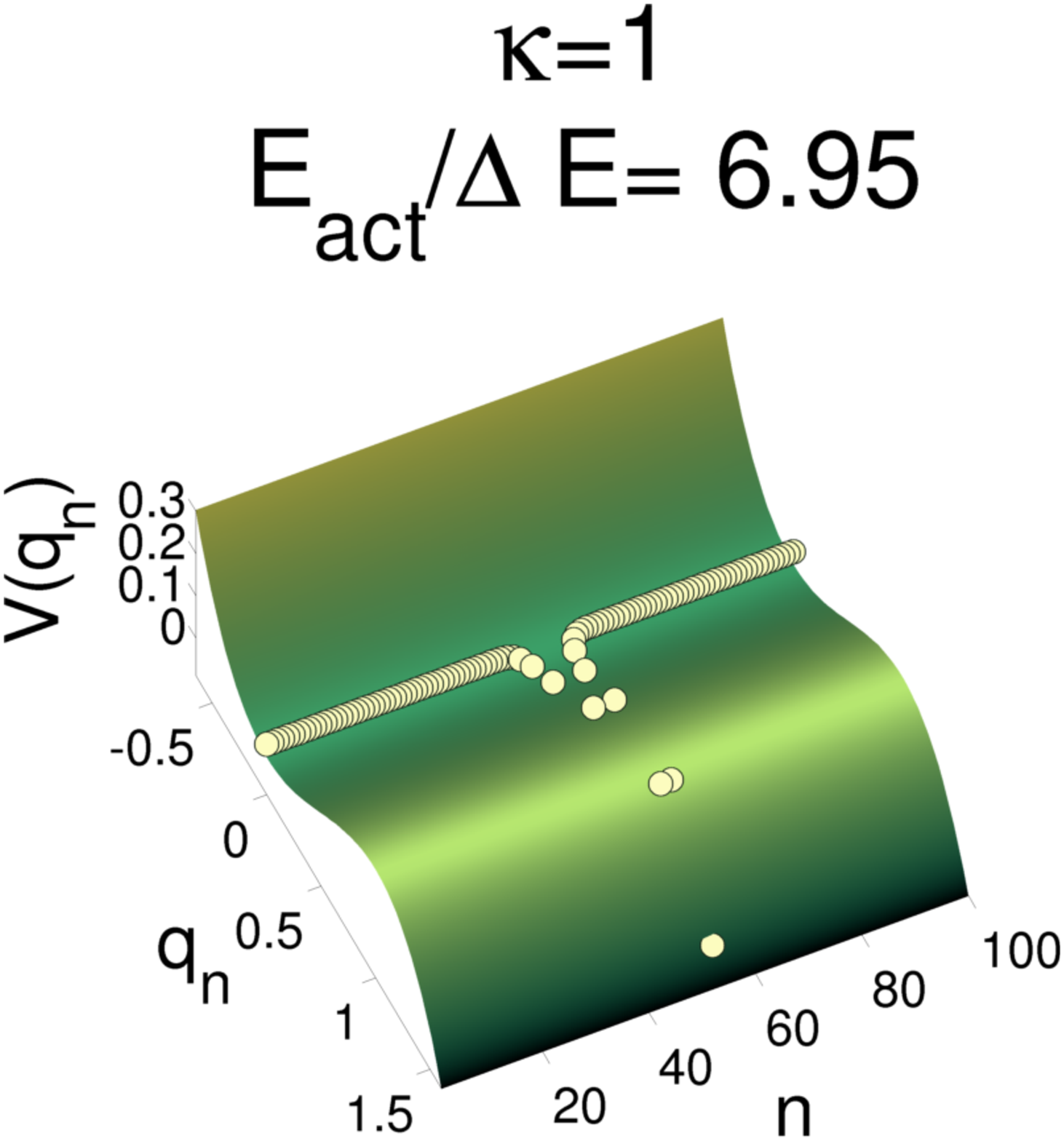}\hfill
\includegraphics[width=0.3\linewidth]{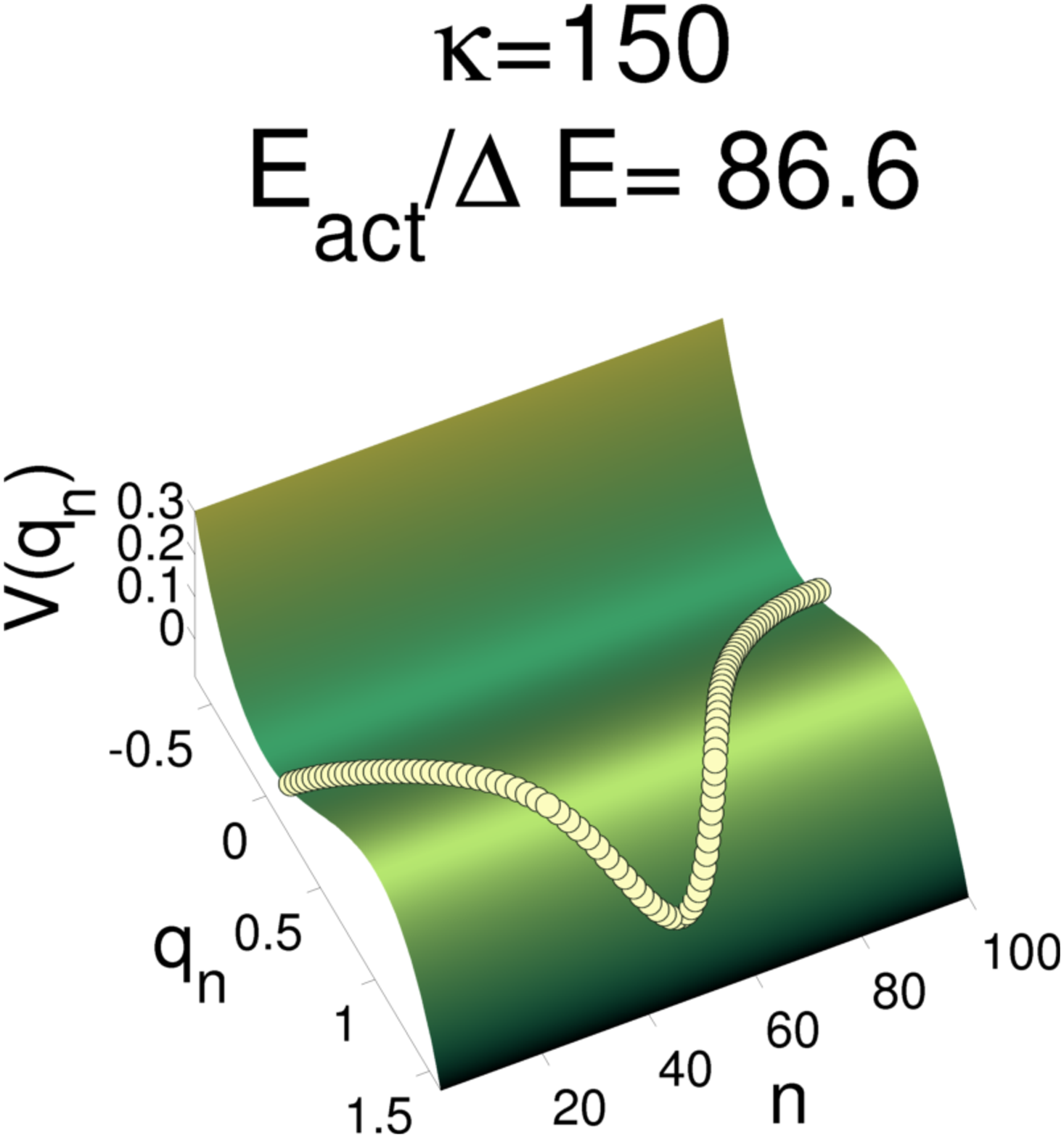}
  \caption{Transition state chain configurations for different values
    of $\kappa$. $N=100$}
  \label{fig:ts:configs}
\end{figure}

%\begin{figure}
%  \centering
%  \epsfig{figure=./images/1d_saddlepoint1,width=0.32\linewidth}
%  \epsfig{figure=./images/1d_saddlepoint2,width=0.32\linewidth}
%  \epsfig{figure=./images/1d_saddlepoint3,width=0.32\linewidth}
%  \caption{Transition state chain configurations for different values
%    of $\kappa$. $N=100$}
%  \label{fig:ts:configs}
%\end{figure}

\begin{figure}
  \centering
  \includegraphics[width=0.5\linewidth]{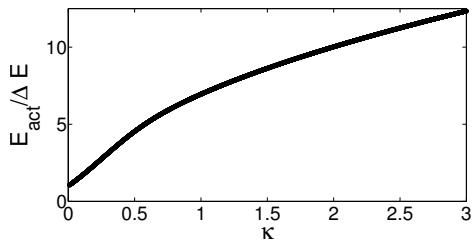}
  \caption{Activation energy (energy of transition state
    configurations), $N=100$}
  \label{fig:ts:E_act}
\end{figure}

In the case of a vanishing coupling strength the oscillators and thus
the equations of motion (\ref{equ_motion_1d}) are no longer coupled.
Consequently, the fixed points of the system consist of all the
configurations where each oscillators is placed either on the maximum
of the potential barrier or the potential valley, $q_n^*=\left\lbrace
  0,1 \right\rbrace$. In this case also the Hessian matrix, $H$,
becomes diagonal and we can directly read off its eigenvalues,
$\lambda^H_n=1-2\,q_n$. Demanding all eigenvalues to be negative
except for one positive the transition states are found to be all the
configurations where all oscillator are positioned in the potential
valley except for one that is placed on the potential barrier. The
according energy reads $E_{\mathrm{act}}(\kappa=0)=\Delta E=1/6$.

In contrast, a very large coupling strength corresponds to a situation
where the chain effectively becomes a single oscillator so that the
transition state refers to a chain configuration where all oscillators
are placed on the maximum of the potential. This can be shown by
taking the limit $\kappa \rightarrow \infty $ in Eq.
(\ref{ts1d:equ_sad_point_1d}). If we want $q_n^*$ to take on values
within a bounded regime we must have $q_{n+1}^*+q_{n-1}^*-2\,q_n^*=0$
in order to satisfy Eq. (\ref{ts1d:equ_sad_point_1d}) in this limit.
In the case of periodic boundary conditions this becomes equivalent to
$q_n^*=q^*$ so that Eq. (\ref{ts1d:equ_sad_point_1d}) becomes
$q^*(1-q^*)=0$. Which of its two roots corresponds to the transition
state becomes clear from the linear stability analysis of Eq.
(\ref{equ_motion_1d}) for this effective one oscillator problem
\[\ddot{q}=-q+{q}^2\approx
-q^*+{q^*}^2+\left(-1+2\,q^*\right)q=\left(-1+2\,q^*\right)q.\] Only
the case $q_n=q^*=1$ is associated with the inherent instability of a transition
state. Accordingly, the transition state energy is found to be
$E_{\mathrm{act}}(\kappa\rightarrow\infty)=N\,\Delta E=N/6$.

The intermediate parameter regime has been evaluated using the dimer
method and the results are represented in Fig. \ref{fig:ts:configs}
and Fig. \ref{ts1d:equ_sad_point_1d}. The maximal amplitude of the
hair pin-like transition state configuration grows with increasing
$\kappa$ until it reaches a critical elongation from which on it
decreases until the entire chain approaches the maximum of the
potential barrier as described above.
  
\section{The formation of breathers}
\label{sec2}
\subsection{Modulational instability of a chain in a nonlinear potential}
The energy that is initially homogeneously distributed along the
entire chain quickly concentrates into local excitations of single
oscillators. This process is governed by the formation of regularly
shaped wave patterns, so-called breathers which are spatially localized and time-periodically varying solutions. 
Their emergence is due to
a modulation instability the mechanism of which applied to our situation is
described later on. We follow \cite{MI_1} and \cite{MI_2} in this
paragraph.

As an approximation for small oscillation amplitudes we can neglect
the nonlinear term in Eq. \ref{equ_motion_1d}. The  resulting equation in  linear
approximation exhibits phonon solutions with frequency $\omega$ and
wave number $k=2\pi\,k_0/N$ (with $k_0\in\mathbb{Z}$ and $-N/2\leq k_o
\leq N/2$) related by the dispersion relation
\begin{align*}
%\label{MI1d1:disperse_linear}
  \omega ^2=1+4\,\kappa\,\sin^2\left(\frac{k}{2}\right)
\end{align*}
We make an Ansatz that only takes into account the first harmonics
(rotating wave approximation)
\begin{align*}%\label{MI1d:Ansatz}
  q_i=F_{1,i}(t)\,e^{-i t}+ F_{0,i}(t)+F_{2,i}(t)\,e^{-2i t}+c.c.
\end{align*}  
The amplitudes of the harmonics are expected to be of a lower order of
magnitude ($| F_{0,i}|\ll| F_{1,i}|$, $| F_{2,i}|\ll| F_{1,i}|$).
Furthermore, we assume our envelope functions to vary slowly
($|\dot{F}_{m,i}|\ll |{F}_{m,i}|$) as well as the phonon band to be
small ($1> 4\kappa$). Within the limits of these assumptions we obtain
a discrete nonlinear Schr\"odinger equation (DNLS) for the amplitudes
of the first harmonic.
\begin{align}\label{MI1d:DNLS}
  2i\dot{F}_{1,i}=\kappa
  \left(\left(F_{1,i-1}+F_{1,i+1}\right)+2F_{1,i}\right)-\frac{10}{3}\,\left|F_{1,i}\right|^2F_{1,i}
\end{align}
We want to study the stability of this equation's plane wave solutions
in the presence of small perturbations $\left|\delta B_i(t)\right|\ll
1$ and $\left|\delta \Psi_i(t)\right|\ll 1$, leading to a new Ansatz
for the envelope function
\begin{align}\label{MI1d:envelope_Ansatz}
  {{F}_{1,n}}^{pert.}=\left(A+\delta B_n(t)\right)
  e^{i\left((k\,n-\Delta\omega\,t)+\delta\Psi_n(t)\right)}.
\end{align}
The perturbations are sufficiently small so that we can expand the
envelope function up to the first order in $\delta$ and neglect all
terms of higher order. Using the Ansatz (\ref{MI1d:envelope_Ansatz})
in Eq. (\ref{MI1d:DNLS}) leads to a complex differential equation for
the perturbation functions $B(t)$ and $\Psi(t)$.  The real and
imaginary part of this equation are independent. Hence, collecting all
terms of first order in $\delta$ results in two linear relations.
\begin{align*}
  -A\,\dot{\Psi}_i&=-\frac{\kappa}{2} \left\lbrace A\,\sin k \left(\Psi_{i-1}-\Psi_{i+1}\right)+\cos k \left(B_{i+1}+B_{i-1}\right)\right\rbrace -\frac{10}{3} \,A^2\, B_i\\
  2 \dot{B}_i&=-\kappa\left\lbrace A\,\cos k \left(\Psi_{i+1}+
      \Psi_{i-1}-2\Psi_i \right)+\sin k
    \left(B_{i+1}-B_{i-1}\right)\right\rbrace
\end{align*} 
Again, the solution to those coupled equations are plane waves
\begin{align*}
  \Psi_n=\Psi ^0 e^{i(Q\,n-\Omega\,t)} \qquad
  B_n=B^0e^{i(Q\,n-\Omega\,t)}
\end{align*}
with the dispersion relation
\begin{align}
  \label{MI1d:pert.-disp_rel}
\left(\Omega-\kappa\,\sin k\,\sin Q\right)^2=
    \kappa\,\cos k\, \sin ^2
    \left(\frac{Q}{2}\right)\left(4\,\kappa\,\cos k \, \sin ^ 2
      \left(\frac{Q}{2}\right)-\frac{20}{3} A^2 \right)
      %\\
  % \left(\Omega-\kappa\,\sin k\,\sin Q\right)^2= \kappa\,\cos k\,
  % \sin ^2 \left(\frac{Q}{2}\right)\left(4\,\kappa\,\cos k \, \sin ^
  %   2 \left(\frac{Q}{2}\right)-\frac{20}{3} A^2 \right)
\end{align}
which describes the stability of the $Q$-mode perturbation on the
$k$-mode carrier wave. $Q$ and $k$ have a $2\,\pi$ periodicity and can
therefore be chosen to be in the first Brillouin zone. Furthermore, we
can restrict the range of $k$ and $Q$: $k,Q\in\lbrace 0 ,\pi \rbrace$,
because negative values correspond to waves with the opposite
direction of propagation.

\begin{figure}
\centering
\begin{minipage}{0.48\linewidth}
\centering
\includegraphics[width=\linewidth]{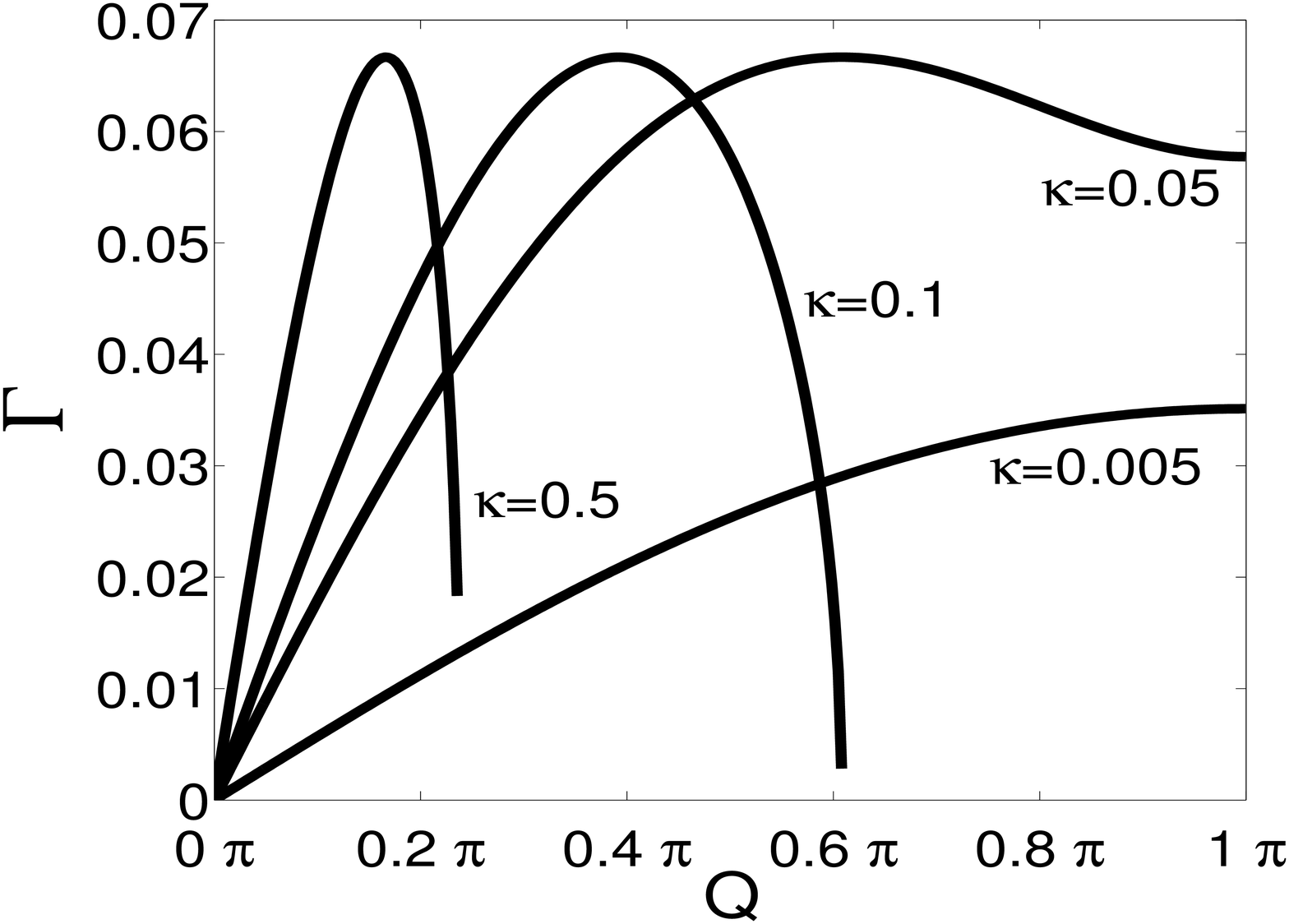}
\caption{Growth rates of unstable Q-modes according to Eq. (\ref{MI1d:growth_rate}) from a k=0 carrier mode with A=0.2}
\label{fig:MI:MI_tau}
\end{minipage}
\hfill
\begin{minipage}{0.48\linewidth}
\centering
\includegraphics[width=\linewidth]{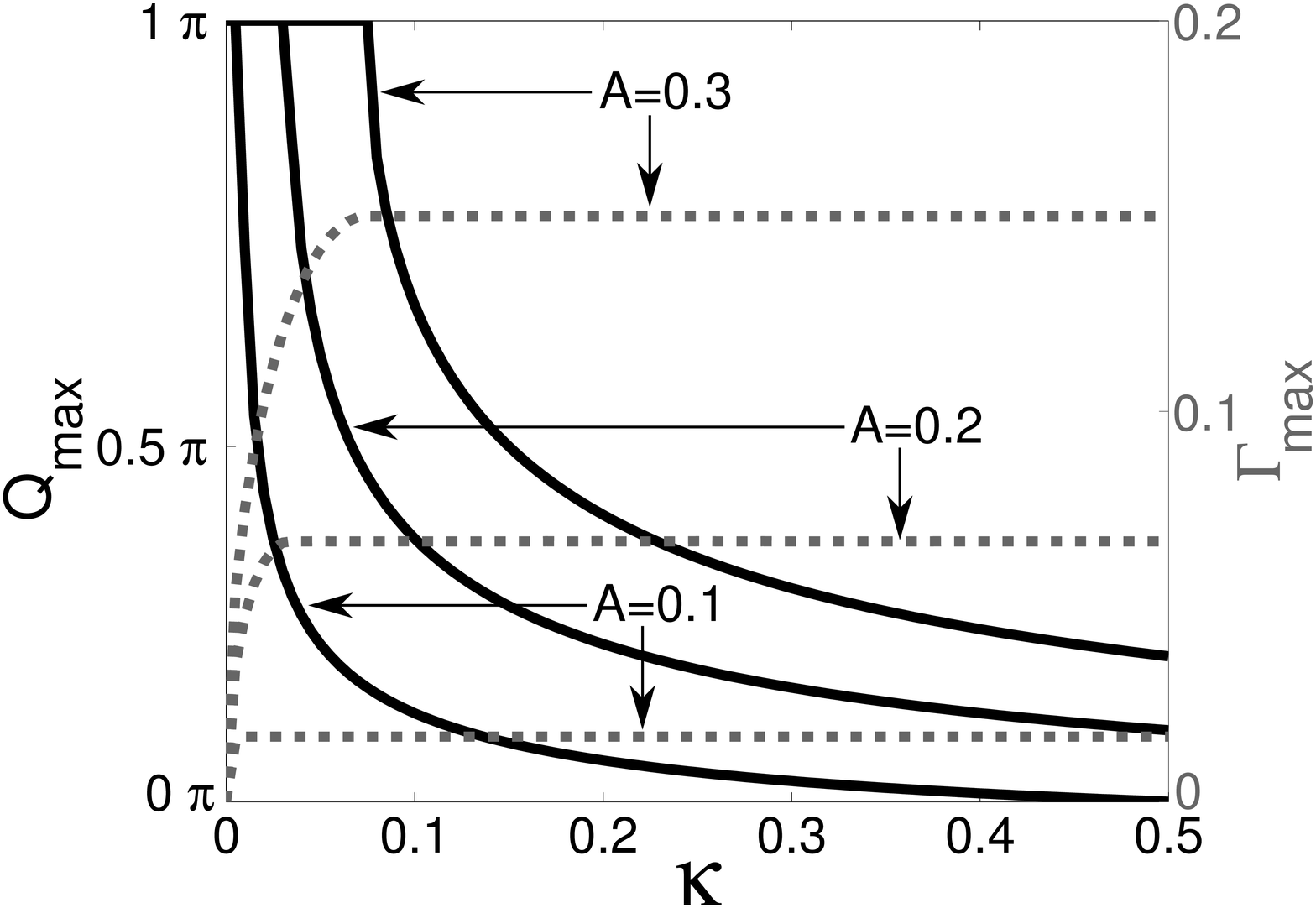}
\caption{Fastest growing modes (solid line) - Eq.
    (\ref{MI1d:Q_max}) - and their growth rates (dashed line), k=0.}
\label{fig:MI:MI_multi}
\end{minipage}
\end{figure}

%\begin{figure}[ht]
%  \centerline{ \minifigure[Growth rates of unstable Q-modes according
%    to Eq. (\ref{MI1d:growth_rate}) from a k=0 carrier mode with
%    A=0.2]
%    {\epsfig{figure=./images/MI_tau,width=2in}\label{fig:MI:MI_tau}}
%    \hspace*{4pt} \minifigure[Fastest growing modes (solid line) - Eq.
%    (\ref{MI1d:Q_max}) - and their growth rates (dashed line), k=0.]
%    {\epsfig{figure=./images/MI_multi,width=2in}\label{fig:MI:MI_multi}}
%  }
%  % \caption{} %rechts:$E=0.072$
%  % \label{fig:MI_multi}% common label
%\end{figure}

The perturbations are stable for $\Omega\in\mathbb{R}$ which is the
case when the right hand side of Eq. (\ref{MI1d:pert.-disp_rel}) is
positive. Therefore, all carrier waves with $k\in\lbrace\pi/2, \pi
\rbrace$ are stable with respect to all perturbation modes. For
$k\in\lbrace 0 , \pi/2\rbrace$ perturbations will grow, provided that
\begin{align}
  \label{MI1d:unstable_Q}
  \cos k \, \sin ^2\left(\frac{Q}{2}\right)&\leq
  \frac{5\,A^2}{3\,\kappa}.
\end{align}
We can then find an according growth rate
\begin{equation}
  \label{MI1d:growth_rate}
  \Gamma (Q)=\left |\operatorname{Im} (\Omega) \right |=\sin\left(\frac{Q}{2}\right)\sqrt{\frac{20}{3}\kappa\,\cos k\,\left({A}^2-\frac{3}{5}\kappa\sin^2 \left(\frac{Q}{2}\right)\cos k\right)}
\end{equation}
which, for the case that $ A^2\leq \frac{6}{5}\kappa \cos k, $ has its
maximum at
\begin{equation}
  \label{MI1d:Q_max}
  Q_\mathrm{max}=2\,\arcsin\sqrt{\frac{5\,A^2}{6\,\kappa\, \cos k}}.
\end{equation}
Otherwise the maximum growth rate is found at $Q=\pi$. The
corresponding growth rates become
\begin{align}\label{MI1d:max_growth_rate}
  \Gamma_\mathrm{max}=
  \begin{cases}\Gamma(Q_\mathrm{max})=\frac{5}{3}A^2&\text{if }A^2\leq \frac{6}{5}\kappa\cos k\\
    \Gamma(\pi)=\sqrt{\frac{20}{3}\kappa\cos k
      (A^2-\frac{3}{5}\kappa\cos k)}<\Gamma(Q_\mathrm{max})&\text{if
    }A^2> \frac{6}{5}\kappa\cos k\\ 
  \end{cases}
\end{align}

We recall from Sect. \ref{1d:1d-chain_model} that our system is
initially prepared in a slightly perturbed $k=0$ mode. This is thus
the only possible carrier wave mode as the amplitudes, $A$, of all
other modes (which scale with the amplitude of the perturbation) are
likely to be too small to generate growing modes -- see inequality
(\ref{MI1d:unstable_Q}) -- or the arising maximal growth rates are
suppressed. Evaluating the growth rate of instabilities on the $k=0$
carrier mode for different values of $\kappa$ (Fig.
\ref{fig:MI:MI_tau}), we find that the modulational instability
becomes more mode selective with increasing $\kappa$. Hence, for large
values of $\kappa$ the only relevant unstable modes are near the
fastest growing mode depicted in Fig. \ref{fig:MI:MI_multi}. In such a
situation we expect the emergence of a regular wave pattern (an array of breathers) that
efficiently localizes energy and thereby enhances the escape of the chain.

\subsection{Optimal coupling}
\label{1d:optimal_coupling}
According to our findings in the previous sectionthe appearing
breather array becomes more regular with increasing $\kappa$ . In particular its prominent mode number, and
therefore the number of breathers, gets smaller, all of which results in an
efficient energy localization. However, an increase in the coupling
strength comes along with an increase in the activation energy -- see
Fig. \ref{fig:ts:E_act} -- which hinders a swift escape for a given
system energy. Therefore, we can expect to find an intermediate
$\kappa$ that optimizes the escape rate.

We can analytically approximate the optimal $\kappa$ by assuming that
the entire system energy is evenly distributed among $N_B$
non-interacting oscillators, where $N_B$ is the number of breathers
related to the prominent wave length, $Q_{\mathrm{max}}$, of the modulational
instability. The ratio of such an oscillator's energy, $E_B$, to the
activation energy can be regarded as a measure of escape efficiency. It is dependent on $Q_{\mathrm{max}}$ which in turn depends on the $k=0$ phonon amplitude $A$ which we
relate to the system energy via $E(A)=N\,V(A)$. Thus we can write  
\begin{align*}
\frac{E_B(\kappa)}{E_{\mathrm{act}}}\propto \frac{1}{Q_{\mathrm{max}}(E,\kappa)\,E_{\mathrm{act}}(\kappa)}.
\end{align*}
This escape efficiency takes on its maximum for the optimal coupling strength, $\kappa
^*$, formally
\begin{align}\label{eq:kappa_reso_analyt}
  \kappa ^*(E) = \underset{ \kappa \in \mathbb{R^+}}{\operatorname{arg\,max}} \, \frac{1}{Q_{\mathrm{max}}(E,\kappa)\,E_{\mathrm{\mathrm{act}}}(\kappa)}.
\end{align}

\begin{figure}
  \centering
   \includegraphics[width=0.65\linewidth]{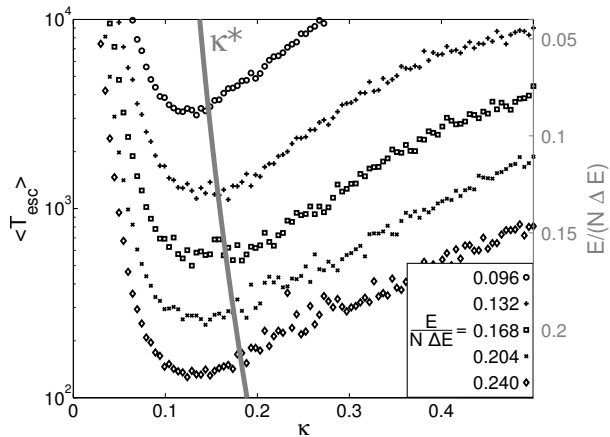}
  \caption{Average escape times (marker symbols) for 500 realizations
    with randomized initial conditions as described in Sec.
    \ref{1d:1d-chain_model}. Parameters: $\Delta q^\mathrm{IC} =0.05,
    \Delta p^\mathrm{IC}=0.05,N=100$. The solid grey line represents
    the analytical approximation for the optimal coupling strength --
    Eq. (\ref{eq:kappa_reso_analyt}) -- for energy values given by the
    right-hand axis.}
  \label{fig:ts:T_esc_optimal_kappa}
\end{figure}

This approximation can now be compared to the numerical evaluation of
average escape times in dependence of $\kappa$, see Fig.
\ref{fig:ts:T_esc_optimal_kappa}. Equation (\ref{equ_motion_1d}) has
been integrated using a fourth order Runge-Kutta scheme. Numerical accuracy was
obtained by ensuring the energy deviation to remain smaller than the
order of $10^{-12}$. The average escape times were determined from $500$
realization of randomized initial conditions at a given energy
according to Sec. \ref{1d:1d-chain_model} for each marker symbol. The
escape time measures the time it takes from  the initialization to
the moment when all oscillators have surpassed the potential barrier.
For all depicted values of $\kappa$ at least $95$ \% of the chain realizations
escaped in the maximal integration time of $5\cdot 10^5$ time units.

\begin{figure}
\centering
\begin{subfigure}[t]{0.48\linewidth}
\includegraphics[width=\linewidth]{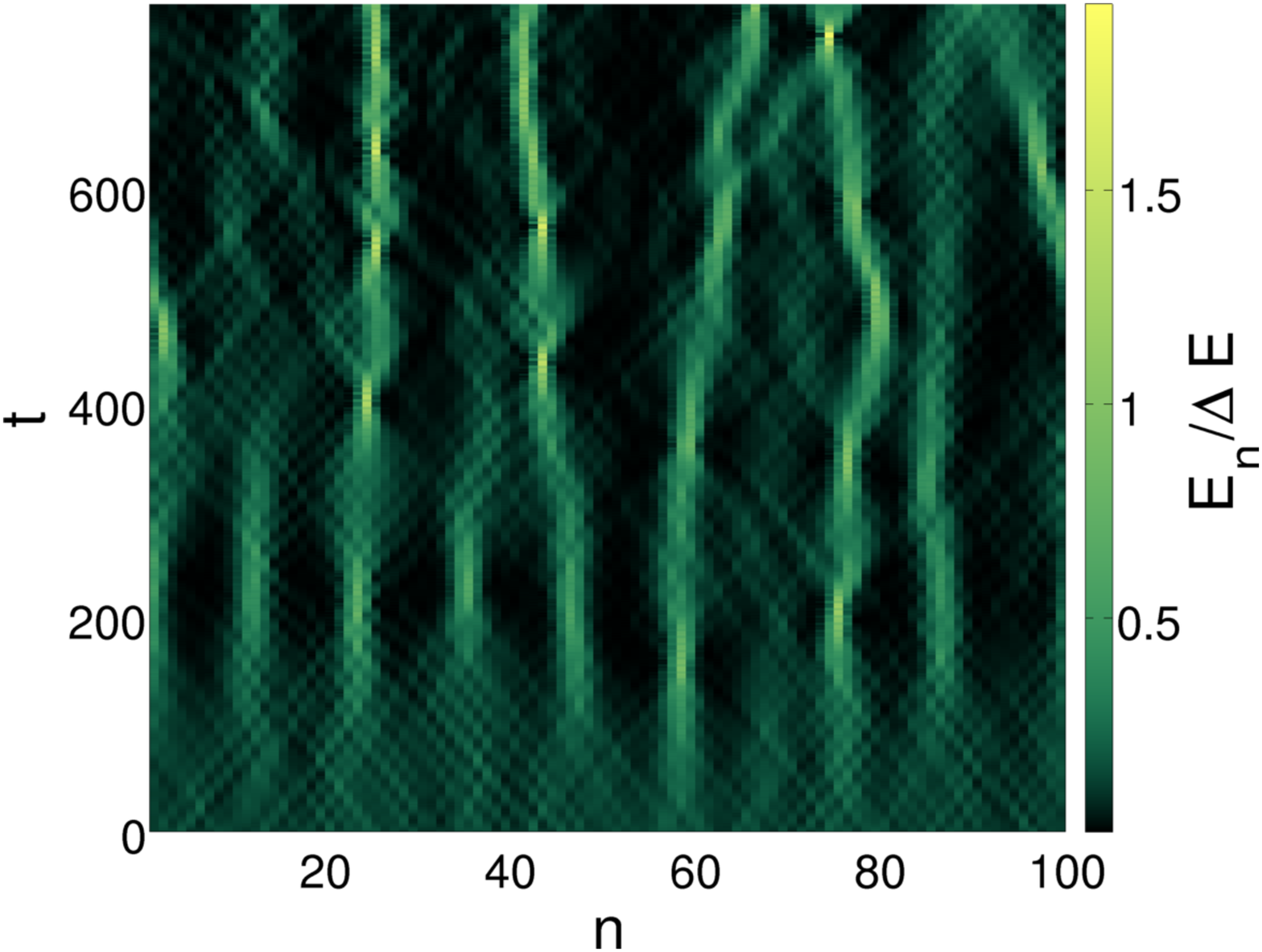}
\caption{$\frac{E}{N\,\Delta E}=0.12$}
\label{fig:ts:energy_evolution_a}
\end{subfigure}
\hfill
\begin{subfigure}[t]{0.48\linewidth}
\includegraphics[width=\linewidth]{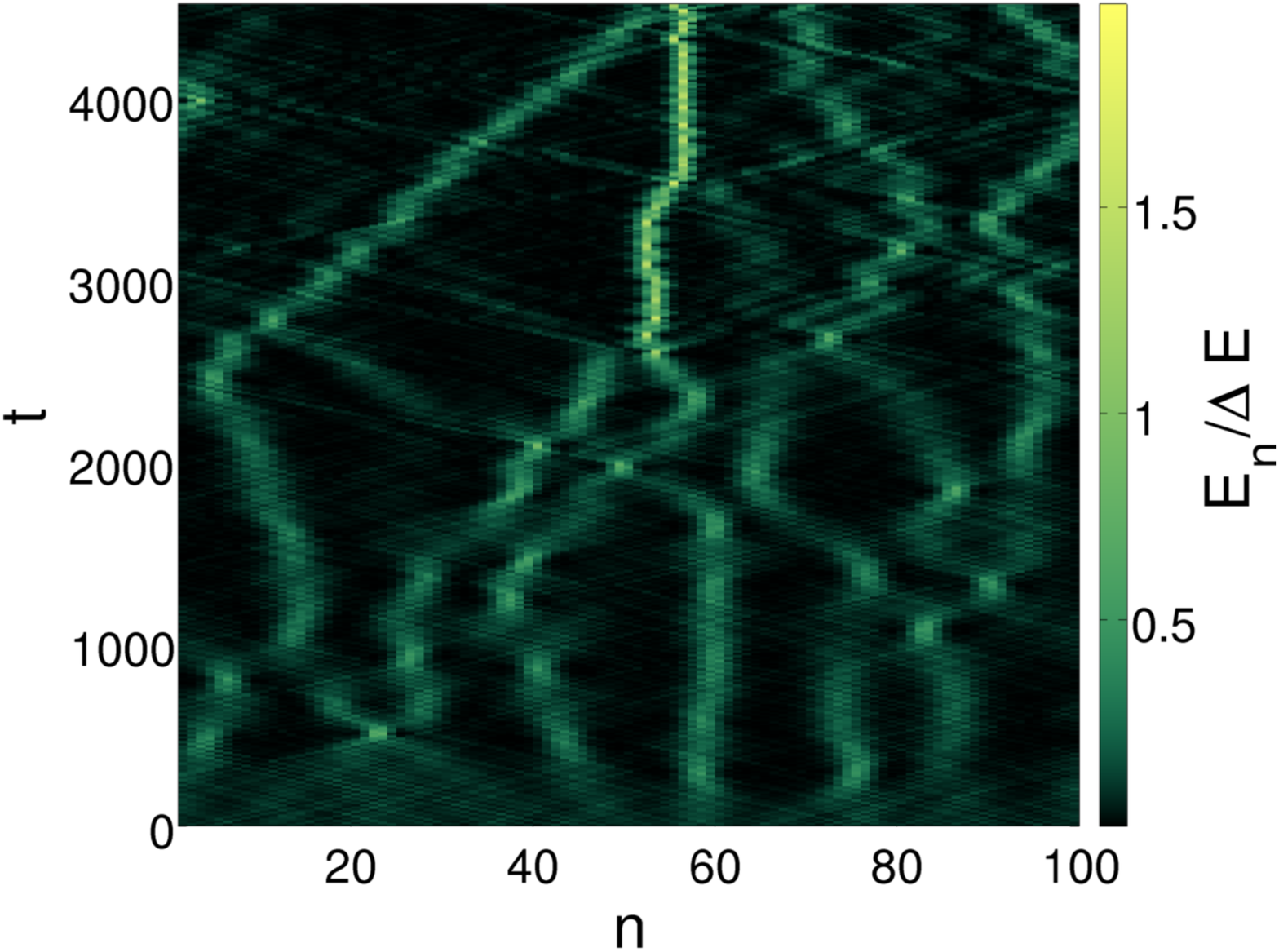}
\caption{$\frac{E}{N\,\Delta E}=0.07$}
\label{fig:ts:energy_evolution_b}
\end{subfigure}
 \caption{Temporal evolution of the energy distribution $E_n(t)$. The
    localization of energy from an initially homogeneous state causes
    in both cases an escape at the end of the depicted time frame.
    While the system energy in Fig. \ref{fig:ts:energy_evolution_a} is
    sufficiently high to let an individual breather of the early
    regular breather array surpass the potential barrier, the lower
    energy in Fig. \ref{fig:ts:energy_evolution_b} necessitates a
    merging of breathers to cause the critical chain elongation.
    Parameters: $\kappa=0.15, \Delta q^\mathrm{IC} =0.05, \Delta
    p^\mathrm{IC} =0.05,N=100$} %rechts:$E=0.072$
  \label{fig:ts:energy_evolution}% common label
\end{figure}

%\begin{figure}[ht]
%  \centerline{ \subfigure[$\frac{E}{N\,\Delta E}=0.12$]
%    {\epsfig{figure=./images/1d_energy_evolution1,width=0.5\linewidth}\label{fig:ts:energy_evolution_a}}
%    \hspace*{4pt} \subfigure[$\frac{E}{N\,\Delta E}=0.07$]
%    {\epsfig{figure=./images/1d_energy_evolution2,width=0.5\linewidth}\label{fig:ts:energy_evolution_b}}
%  }
%  \caption{Temporal evolution of the energy distribution $E_n(t)$. The
%    localization of energy from an initially homogeneous state causes
%    in both cases an escape at the end of the depicted time frame.
%    While the system energy in Fig. \ref{fig:ts:energy_evolution_a} is
%    sufficiently high to let an individual breather of the early
%    regular breather array surpass the potential barrier, the lower
%    energy in Fig. \ref{fig:ts:energy_evolution_b} necessitates a
%    merging of breathers to cause the critical chain elongation.
%    Parameters: $\kappa=0.15, \Delta q^\mathrm{IC} =0.05, \Delta
%    p^\mathrm{IC} =0.05,N=100$} %rechts:$E=0.072$
%  \label{fig:ts:energy_evolution}% common label
%\end{figure}

Figure \ref{fig:ts:T_esc_optimal_kappa} clearly shows the predicted
resonance behaviour and also reveals a fairly good accordance of the
analytical approximation of the optimal $\kappa$ with the simulation
results. This seems to verify our initial assumption of a regular
breather array that fully concentrates the energy into single oscillators.
But this reasoning fails to explain the pronounced variation of the average
escape times (ranging over several orders of magnitude) for different
energies. Additionally, the equipartition of a low system energy will
not allocate single breathers with an energy sufficient to trigger an
escape event. E.g. for $\kappa=0.15$ and $E/(N\,\Delta E)=0.1$ we
expect an array with ten or more breathers so that each one could only
hold an energy $E/N_B<\Delta E$. Nevertheless, an escape takes place,
eventually. This implies a further concentration of energy beyond the
initial creation of the breather array. In order to study this
process, we look at the snapshots of the energy distribution
\begin{align*}
  E_n=\frac{p_n^2}{2}+V(q_n)+\frac{\kappa}{4}\left\lbrace
    (q_n-q_{n+1})^2+ (q_{n-1}-q_{n})^2\right\rbrace,
\end{align*} 
In Fig. \ref{fig:ts:energy_evolution} $E_n$ has been tracked in time
(upwards) for two exemplary cases. Energy is localized in both cases
starting from an initially homogeneous state. In Fig.
\ref{fig:ts:energy_evolution_a} we see the appearance of a regular
breather pattern. Every breather concentrates enough energy to certain
oscillators in order to trigger an escape. In contrast, the lower system energy in Fig. \ref{fig:ts:energy_evolution_b} does not allow for a direct escape of the initial breathers. Instead, breathers
start an erratic movement. After an inelastic interaction they merge
and can thereby eventually result in a configuration exceeding the critical
chain elongation, see also \cite{Peyrard1998_localization}. However,
this secondary process is slow compared to the (direct) breather formation
which explains the different orders of magnitude of the escape
times scale in Fig.  \ref{fig:ts:T_esc_optimal_kappa}.

\section{Thermally activated escape supported by brea\-thers}
\label{sec3}
In the previous sections we have been concerned with the deterministic
chain dynamics leading to an escape event. In this section we study
how a thermal bath with temperature $T$ will modify the transition over the barrier. For
this purpose we consider the associated Langevin equation,
\begin{align}
  \label{1d_Langevin_eq}
  \ddot{q_n}+q_n-{q_n}^2-\kappa
  \left(q_{n+1}+q_{n-1}-2\,q_n\right)+\gamma \dot{q_n}+\xi_n(t)=0,
\end{align}
with the friction parameter $\gamma$ and a Gaussian white noise term
$\xi_n(t)$. In order to be able to compare the deterministic situation
to the thermally activated setting, the associated conserved energy
$E$ in the Hamiltonian case and the average energy $\overline{E}$
transferred from the bath need to be equal. The latter is governed by
the correlation function of the noisy force $\xi(t)$. Its permanent
variation yields source of energy for the chain which is balanced by
the dissipative friction forces.

The transferred energy is defined if the autocorrelations functions of
the noise sources scale as
\begin{align}
\label{fdt}
\left\langle \xi_n(t) \xi_{n^\prime}(t^\prime) \right\rangle=2\,\gamma
\,\overline{E}/N \,\delta_{n,n^\prime}\,\delta(t-t^\prime)\,.
\end{align}
This relation, known as fluctuation dissipation theorem, implies that
the mean energy of all particles is given by $\overline{E}$. If
expressed by the bath temperature, every particle gets in averrage
$k_B\,T$, {\it i.e.} $\overline{E} = N\, k_B\, T$ with $k_B$ being the Boltzmann
constant.

In numerical simulations of the Langevin equation with fulfilled
relation (\ref{fdt}), we have assured that the full average energy
converges to $\overline{E}$ also for the transient state of the
transition. Initially after the chain has relaxed to a stationary
situation around the metastable minimum of the potential $V(q_{min})$,
the oscillators obey the canonic distribution in phase space near to
this minimum.

\begin{figure}
\centering
\begin{subfigure}[t]{0.49\linewidth}
\includegraphics[width=\linewidth]{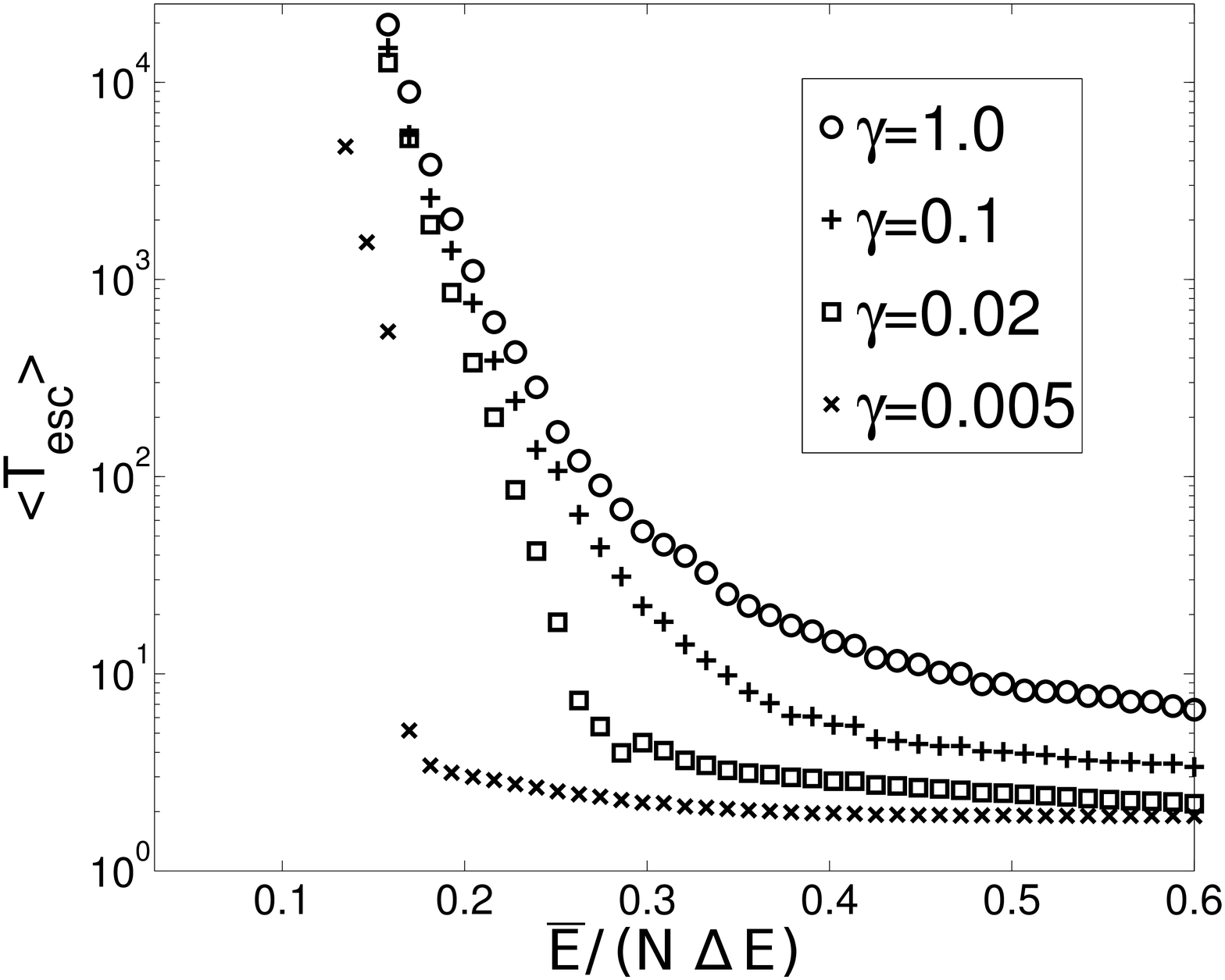}
\caption{Thermally activated setting -- Eq. (\ref{1d_Langevin_eq})}
\label{fig:comp_plot1}
\end{subfigure}
\hfill
\begin{subfigure}[t]{0.49\linewidth}
\includegraphics[width=\linewidth]{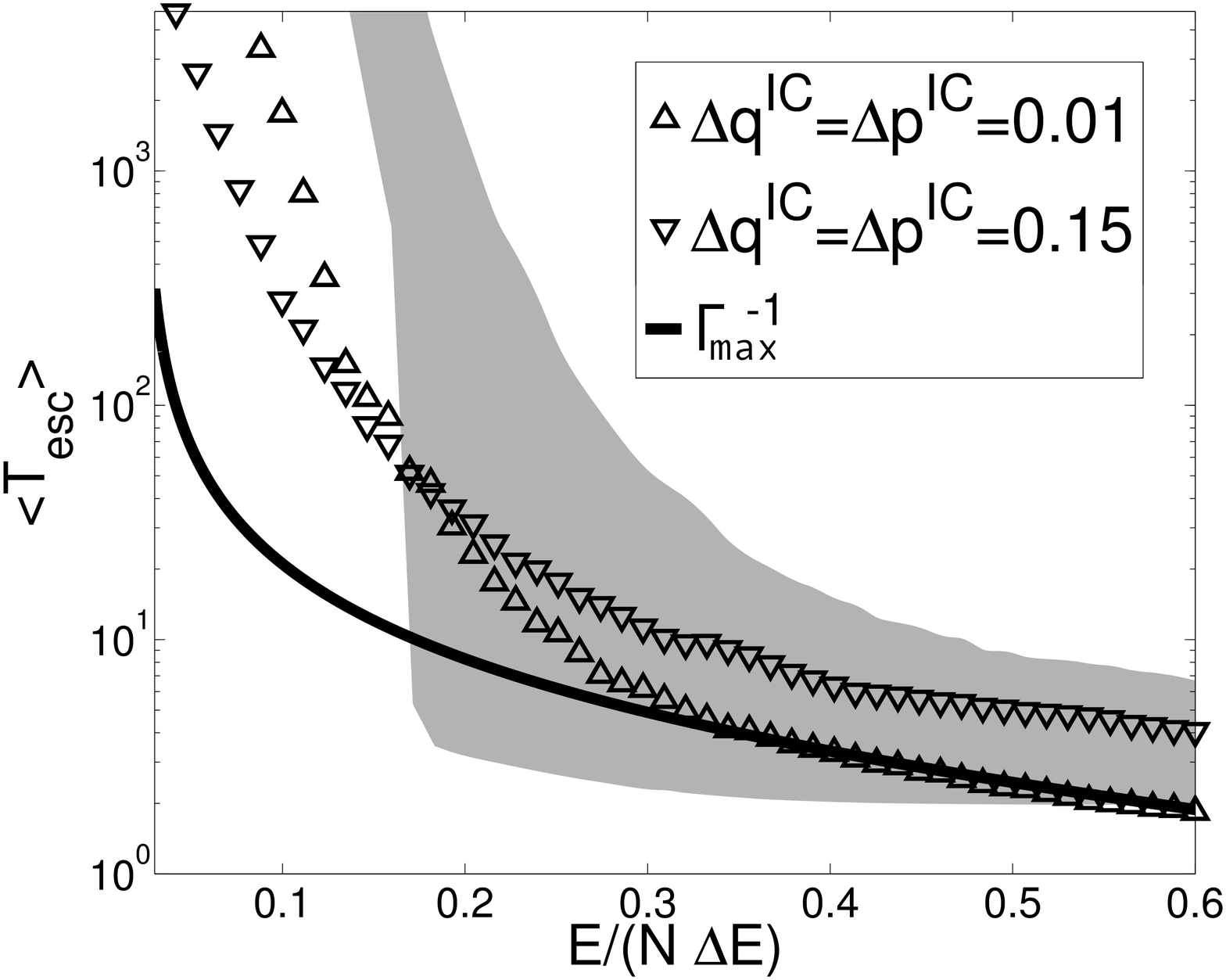}
\caption{Deterministic setting -- Eq. (\ref{equ_motion_1d})}
\label{fig:comp_plot2}
\end{subfigure}
\caption{Average escape times in the thermally activated case for
    different values of the friction constant and their comparison to
    the deterministic setting for 500 realizations each. The grey area
    in Fig. \ref{fig:comp_plot2} sketches the average escape times for
    the thermally activated case for $0.005 < \gamma < 1.0$ as shown
    in Fig. \ref{fig:comp_plot1} where as the symbols shows results
    from the deterministic set-up with initial conditions given in the
    inset. Parameters: $\kappa=0.15, N=100$}
  \label{fig:comp_plot}
\end{figure}

%\begin{figure}[ht]
%  \centerline{ \subfigure[Thermally activated setting -- eq.
%    (\ref{1d_Langevin_eq})]
%    {\epsfig{figure=./images/comp_plot1,width=0.5\linewidth}\label{fig:comp_plot1}}
%    \hspace*{4pt} \subfigure[Deterministic setting -- Eq.
%    (\ref{equ_motion_1d}). ]
%    {\epsfig{figure=./images/comp_plot2,width=0.5\linewidth}\label{fig:comp_plot2}}
%  }
%  \caption{Average escape times in the thermally activated case for
%    different values of the friction constant and their comparison to
%    the deterministic setting for 500 realizations each. The grey area
%    in Fig. \ref{fig:comp_plot2} sketches the average escape times for
%    the thermally activated case for $0.005 < \gamma < 1.0$ as shown
%    in Fig. \ref{fig:comp_plot1} where as the symbols shows results
%    from the deterministic set-up with initial conditions given in the
%    inset. Parameters: $\kappa=0.15, N=100$}
%  \label{fig:comp_plot}% common label
%\end{figure}

We measure average escape times for the system described by Eq.
(\ref{1d_Langevin_eq}). The latter is numerically integrated using an
Euler scheme, again with a maximal integration time of $5\cdot 10^5$ time units. The system is initialized with
all oscillators set to the minimum of the potential and zero momenta.
We then let the system thermalize until its energy reaches
$\overline{E}$ for the first time. The time from this moment until all
oscillators have surpassed the potential's maximum counts as the
escape time.

We study the system for the optimal coupling constant, $\kappa=0.15$,
as described in Sec. \ref{1d:optimal_coupling}. Figure
\ref{fig:comp_plot1} shows the average escape times of 500
realizations for different values of the friction constant in dependence of $\overline{E}$. Figure
\ref{fig:comp_plot2} compares these times (depicted as the grey
surface) to the according average escape times of the deterministic
system. It additionally shows the characteristic time constant for the formation of breathers, $\Gamma_{\mathrm{max}}^{-1}$, taken from Eq. (\ref{MI1d:max_growth_rate}), where again we relate the $k=0$ phonon amplitude, $A$, to the system energy via $E(A)=N\,V(A)$.

Especially for smaller energies the deterministic escape is considerably faster than the thermally activated one. Notably for $E/(N\,\Delta E)<0.1$ and quite contrary to the deterministic setting, escape events are practically absent during our simulations time in the thermal case. For larger
energy values this picture can change to a higher efficiency of the thermal escape process when damping is weak. Also two deterministic settings with different magnitudes of the random initial perturbations have a converse behaviour for low and high energies. 

\begin{figure}
\centering
\begin{subfigure}[t]{0.48\linewidth}
\includegraphics[width=\linewidth]{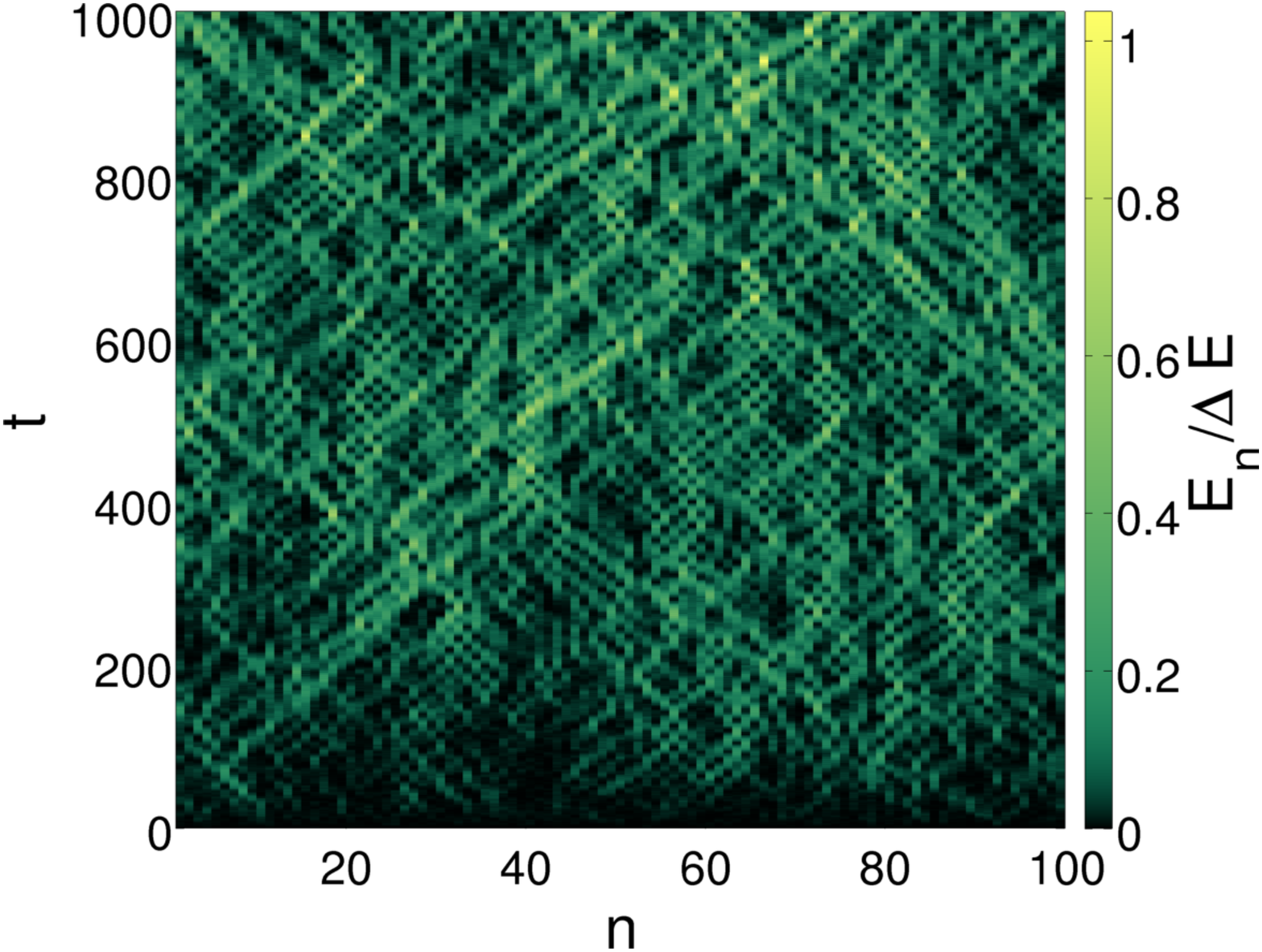}
\caption{$\gamma=0.005$}
\label{fig:1d:E_evo_thermal_1}
\end{subfigure}
\hfill
\begin{subfigure}[t]{0.48\linewidth}
\includegraphics[width=\linewidth]{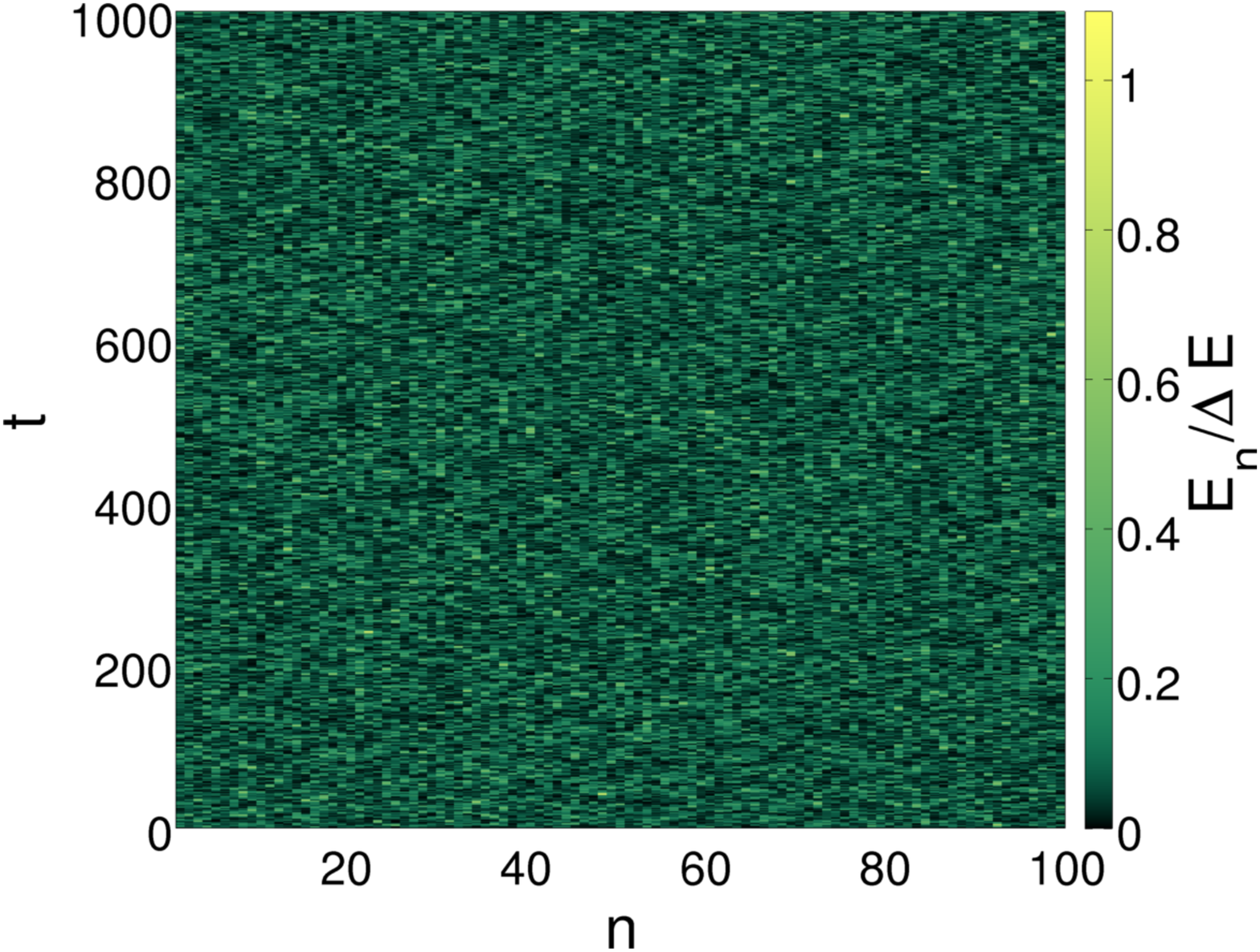}
\caption{$\gamma=1.0$}
\label{fig:1d:E_evo_thermal_2}
\end{subfigure}
  \caption{Temporal evolution of the energy distribution $E_n(t)$ in
    the thermally activated case. A smaller friction constant leads to
    a higher degree of energy localization. Parameters:
    $\overline{E}/(N\,\Delta E)=0.15, \quad \kappa=0.15$.}
  \label{fig:1d:E_evo_thermal}
\end{figure}

%\begin{figure}[ht]
%  \begin{center}
%    \subfigure[$\gamma=0.005$]
%    {\epsfig{figure=./images/E_evol_thermal_gam_i005,width=0.49\linewidth}\label{fig:1d:E_evo_thermal_1}}
%    \hfill \subfigure[$\gamma=1.0$]
%    {\epsfig{figure=./images/E_evol_thermal_gam_1i,width=0.49\linewidth}\label{fig:1d:E_evo_thermal_2}}
%  \end{center}
%  \caption{Temporal evolution of the energy distribution $E_n(t)$ in
%    the thermally activated case. A smaller friction constant leads to
%    a higher degree of energy localization. Parameters:
%    $\overline{E}/(N\,\Delta E)=0.15, \quad \kappa=0.15$.}
%  \label{fig:1d:E_evo_thermal}% common label
%\end{figure}
%
%

How can this result be explained by observations of the chain
dynamics?  For small values of $\gamma$ the system approaches the
deterministic setting. This entails an observable tendency towards a
more localized energy distribution as seen in comparing Fig. \ref{fig:1d:E_evo_thermal_1} and Fig. \ref{fig:1d:E_evo_thermal_2}. The relaxation time of the chain scales with the inverse of the damping constant. Correspondingly, the life times of local excitations grow with decreasing $\gamma$. The outcome is a more heterogeneous energetic structure where thermal fluctuations are more likely to cause critical chain elongations. This explains the faster escape comparing small with large damping constants.
But even in the case of very small $\gamma$ the relaxation time is still much shorter 
than the time needed for the coalescence of multiple breathers (which is of the order of
several hundred time units, see Fig. \ref{fig:ts:energy_evolution_b}). Therefore, the long term  cumulative concentration of energy, as described in \ref{1d:optimal_coupling}, is generally inhibited in the thermal case. This explains the virtual impossibility of a thermal escape for small energies.

In the opposite case of higher energies the deterministic escape is mostly proceeded by initial breathers. The formation time of the initial breather array can be estimated by the inverse of the maximal growth rate, $\Gamma_{\mathrm{max}}$ from Eq. \ref{MI1d:max_growth_rate}. Figure \ref{fig:comp_plot2} shows the expected convergence of $\Gamma_{\mathrm{max}}^{-1}$ to the deterministic escape times for large energies. We recall from Sec. \ref{sec2} that $\Gamma_{\mathrm{max}}$ was determined starting with a linear expansion in the perturbations. We believe that this explains the better match in the case of smaller initial perturbations. The divergence of $\Gamma_{\mathrm{max}}^{-1}$ with the average escape times for small energies again gives evidence to the fact that the initial breather array does not induce an escape and other mechanisms are needed.

In the thermally activated case the large energy setting holds two characteristic scenarios. 
The mobility of the breathers becomes amplified if noise acts. If the breathers possess longer life times, {\it i.e.}  for smaller damping, a few (usually not more the two) breathers can temporarily     merge and thus approach a critical chain elongation. Starting with a heterogeneous energy structure after thermalization, this then leads to smaller escape times compared to the deterministic setting that first has to re-allocate energy from an initially
homogeneous state. Oppositely for
stronger damping, the life times of breather is too short and the energy distribution is mostly homogeneous (see Fig. \ref{fig:1d:E_evo_thermal_2}). The escapes then relies entirely on rare, large enough spontaneous
fluctuation of the noise term and the average escape times generally become comparatively large. 

Finally, we want to examine the converse behaviour for different magnitudes of the initial perturbations in the deterministic case. As the initial conditions for smaller perturbations are closer to the $k=0$ phonon mode the initial breather array emerges more quickly so that the escape times are smaller when the energy is high enough for initial breathers to ignite the escape. Contrarily, stronger perturbations lead to a higher breather mobility which accelerates the breather coalescence so that the escape times become smaller when low energy necessitates coalescence.

\section{Summary} 
\label{sec:4}

We have studied various examples of collective escape processes in
many-particle systems. First we started with a noise-free escape in a
chain of coupled oscillators evolving in a metastable potential. The second part of this work examined a thermally activated
escape. To this end the original system has been
augmented by a linear friction term and Gaussian thermal, white noise
of vanishing mean satisfying the fluctuation dissipation theorem.

While a thermally activated escape becomes virtually impossible at low system energy, the deterministic system remains capable of efficiently crossing the potential barrier. In more detail, the deterministic dynamics of interacting chain units
leads to the formation of breather solutions localizing energy such
that the chain passes through a transition state and crosses the
potential barrier. In particular at low system energy, the
interaction between several formed breathers resulting in their
coalescence eventually enables the chain to accumulate sufficient
energy to overcome the potential barrier. Interestingly, in the thermally activated setting with a sufficiently weak damping, the breather formation and thermal
fluctuation cooperate to localize energy which accomplishes effective
barrier crossings.  

We underline that the dynamics of interacting particles exhibiting
collective behaviour such as breather formation and their interaction
has a huge impact on the escape and activation dynamics in such
many-body systems. Hence, our study intends to offer new perspectives
on the understanding of such collective escape processes.

\section{Acknowledgments}
L. Schimansky-Geier thanks for support from IRTG 1640 of the
Deutsche Forschungsgemeinschaft. The authors acknowledge previous co-authors for a successful cooperation.\\

Preprint of an article published in the book ``First-Passage Phenomena and Their Applications'': pp. 554-570, May 2014;  doi: 10.1142/9789814590297{\_}0022 \textcopyright World Scientific Publishing Company.

\end{document}